\documentclass{article}
\usepackage{multicol}
\usepackage{ragged2e}
\usepackage{graphicx,setspace}
\usepackage{natbib}
\usepackage{subfig}
\usepackage{fancyhdr,amsmath,amssymb, amsthm,mathtools}
\usepackage{etoolbox}
\usepackage[ruled,norelsize,vlined]{algorithm2e}

\usepackage[colorlinks,citecolor=blue,urlcolor=blue]{hyperref}
\usepackage{booktabs}
\usepackage{tikz}
\usetikzlibrary{matrix,positioning}
\usetikzlibrary{fit,positioning,calc}
\usepackage{scrextend}
\usepackage{hyperref}
\usepackage{chngcntr}
\usepackage{apptools}
\usepackage{bbm}
\usepackage{ulem}
\usepackage{caption}
\usepackage[T1]{fontenc}
\usepackage{tgtermes}
\usepackage[small]{titlesec}

\newcommand{\bx}{ {\bf x} }
\newcommand{\bof}{ {\bf f} }
\newcommand{\by}{ {\bf y} }
\newcommand{\ba}{ {\bf a} }
\newcommand{\bb}{ {\bf b} }
\newcommand{\bq}{ {\bf q} }
\newcommand{\bk}{ {\bf k} }
\newcommand{\bs}{ {\bf s} }
\newcommand{\bz}{ {\bf z} }
\newcommand{\bX}{ {\bf X} }
\newcommand{\bH}{ {\bf H} }
\newcommand{\bI}{ {\bf I} }
\newcommand{\bD}{ {\bf D} }

\newcommand{\bbeta}{ {\boldsymbol \beta} }

\newcommand{\bzero}{ {\bf 0} }
\newcommand{\bv}{ {\bf v} }
\newcommand{\bL}{ {\bf L} }
\newcommand{\bu}{ {\bf u} }

\newcommand{\bU}{ {\bf U} }
\newcommand{\balpha}{ {\boldsymbol \alpha} }
\newcommand{\bSigma}{ {\boldsymbol \Sigma} }
\newcommand{\bmu}{ {\boldsymbol \mu} }

\newcommand{\bxi}{ {\boldsymbol \xi} }
\newcommand{\bOmega}{ {\boldsymbol \Omega} }
\newcommand{\bPsi}{ {\boldsymbol \Psi} }
\newcommand{\boeta}{ {\boldsymbol \eta }}

\newcommand{\bomega}{ {\boldsymbol \omega} }

\newcommand{\bw}{ {\bf w} }
\newcommand{\pr}{ \mbox{pr} }

\newtheorem{Proposition}{Proposition}

\newtheorem{Lemma}{Lemma}

\begin{document}

\title{Scalable Computation of Predictive Probabilities in Probit Models with Gaussian Process Priors}
	\author{Jian Cao\footnote{Statistics Program, King Abdullah University of Science and Technology, Saudi Arabia, e-mail: jian.cao@tamu.edu}, \ Daniele Durante\footnote{Department of Decision Sciences and Bocconi Institute for Data Science and Analytics, Bocconi University, Italy, e-mail: daniele.durante@unibocconi.it}  \ and Marc G. Genton\footnote{Statistics Program, King Abdullah University of Science and Technology, Saudi Arabia, e-mail: marc.genton@kaust.edu.sa}}
	\date{}
	\maketitle

\vspace{-3pt}
\begin{abstract}
Predictive models for binary data are fundamental in various fields, and the growing complexity of modern applications has motivated several flexible specifications for modeling the relationship between the observed predictors and the binary responses. A widely-implemented solution is to express the probability parameter via a probit mapping of a Gaussian process indexed by predictors. However, unlike for continuous settings, there is a lack of closed-form results for predictive distributions in binary models with Gaussian process priors. Markov chain Monte Carlo methods and approximation strategies provide common solutions to this problem, but state-of-the-art algorithms are either computationally intractable or inaccurate in moderate-to-high dimensions. In this article, we aim to cover this gap by deriving closed-form expressions for the predictive probabilities in probit Gaussian processes that rely either on cumulative distribution functions of multivariate Gaussians or on functionals of multivariate truncated normals. To evaluate these quantities  we develop novel scalable solutions based on tile-low-rank Monte Carlo methods for computing multivariate Gaussian probabilities, and on mean-field variational approximations of multivariate truncated normals. Closed-form expressions for the marginal likelihood and for the posterior distribution of the Gaussian process are also discussed. As shown in simulated and real-world empirical studies, the proposed methods scale to dimensions where state-of-the-art solutions are impractical.   
\end{abstract}
\noindent%

\begin{multicols}{2}

\section{Introduction}\label{sec.1}
There is an increasing demand in various fields of application for flexible models that can accurately characterize complex relations among a vector of binary response data $\by=(y_1, \ldots, y_n)^{\intercal}$ and a set of predictors $\bX=(\bx_{1}, \ldots, \bx_{n})^{\intercal}$, where $y_i \in \{0; 1\}$, whereas $\bx_{i}=(x_{i1}, \ldots, x_{iq})^{\intercal} \in \Bbb{R}^q$, for  every unit $i=1, \ldots, n$. Common solutions address this goal by replacing the  linear predictor $\bX\bbeta=(\bx_1^{\intercal}\bbeta, \ldots, \bx_n^{\intercal}\bbeta)^{\intercal} \in  \Bbb{R}^n$ within the generalized linear model for $\by$ \citep{nelder1972} with a more flexible vector 
\begin{equation*}
\bof(\bX)=\left(f(\bx_1), \ldots,f(\bx_n)\right)^{\intercal} \in  \Bbb{R}^n, 
\end{equation*}
which accounts for complex non-linear relations between the response and the predictors, thus enhancing predictive power. Notable examples of this approach within the Bayesian setting define $\bof(\bX)$ via additive trees \citep{chipman_2010}, Bayesian P--splines \citep{brezger2006} and Gaussian processes (GP) \citep{ras_2006}, among others. 

Motivated by the success of GP for classification  \citep[][]{neal1999, opper2000, de2005, chu2005, kuss2005, girolami2006,ras_2006,choudhuri2007,riihimaki2013}, we aim at deriving improved methods to evaluate  predictive probabilities within this class of models under the probit link. Following the standard practice, we assume that  $y_i$, for $i=1, \ldots, n$, are conditionally independent realizations from Bernoulli variables with probabilities $\Phi(f(\bx_i))=\pr(y_i=1 \mid f(\bx_i))$, $i=1, \ldots, n$, where $\Phi(f(\bx))$ is the cumulative distribution function of a standard Gaussian  evaluated at $f(\bx)$, whereas $f(\bx)$   is assigned a GP prior with mean function $m(\bx)=\mathbb{E}(f(\bx))$ and covariance kernel $K(\bx,\bx')=\mathbb{E}[(f(\bx)-m(\bx))(f(\bx')-m(\bx'))]$. In routine implementations \citep[e.g.,][]{kuss2005,ras_2006}, $K(\bx,\bx')$ denotes a pre-specified function indexed by a low-dimensional vector of hyperparameters $\balpha \in  \Bbb{R}^d$, where $d \in \{1;2;3\}$ in commonly implemented covariance functions  \citep[][Ch.\ 4.2]{ras_2006}. These quantities can be either fixed to default values by inheriting  guidelines from Bayesian regression for binary data \citep[][]{gelman_2008,chopin_2017}, or can be estimated leveraging information from observed data via direct maximization of the marginal likelihood  \citep[e.g.,][]{kuss2005,ras_2006}; see Section~\ref{sec.5} for a discussion on estimation of $\balpha$ in large $d$ settings. The mean function $m(\bx)$ is, instead, commonly set equal to $0$, or is assigned a further layer of hierarchy which typically specifies $m(\bx)$ via a linear combination $\bx^{\intercal}\bbeta$ of the predictors $\bx$, where $\bbeta$ denotes a $q$-dimensional vector of coefficients generally assumed to have independent Gaussian priors $\mbox{N}(0, \delta^2)$ \citep[e.g.,][Ch.\ 2.7]{ras_2006}. Although estimation and uncertainty quantification for $\bbeta$ can be of interest, the key aim of this article is to improve predictive inference in probit GPs. Such a goal is in line with the general focus of GP literature that often employs Gaussian process representations to improve predictive performance relative to classical linear regression models \citep[e.g.,][]{kuss2005,ras_2006,girolami2006,nic2008,riihimaki2013}. Consistent with this goal, when $\bx^{\intercal}\bbeta$ enters the GP mean function, we follow  \citet[][Ch.\ 2.7]{ras_2006} by marginalizing out $\bbeta$ and evaluating predictive probabilities under the induced GP prior for $f(\bx)$, with mean function equal to $0$ and covariance kernel given by $K(\bx,\bx')+\delta^2 \bx^{\intercal} \bx'$. As discussed in \citet[][Ch.\ 2.7]{ras_2006}, this updated kernel formally allows to fully exploit possible  linear relationships among the response and covariates in predictive inference.

Leveraging  basic GP properties \citep[][]{ras_2006} and assuming, without any loss of generality, no overlap in $\bx_1, \ldots, \bx_n$, the aforementioned probit Gaussian process models can be generally expressed as
\begin{equation}
\begin{split}
&p(\by \mid \bof(\bX))= \prod\nolimits_{i=1}^n \Phi(f(\bx_i))^{y_i}[1-\Phi(f(\bx_i)) ]^{1-y_i}, \\
&  p(\bof(\bX))= \phi_n(\bof(\bX)-\bxi; \bOmega),
\end{split}
\label{eq1}
\end{equation}
where $\phi_n(\bof(\bX)-\bxi; \bOmega)$ denotes the density function of a multivariate Gaussian distribution $\mbox{N}_n(\bxi, \bOmega)$ for $\bof(\bX)$, with mean vector $\bxi=(m(\bx_1), \ldots, m(\bx_n))^{\intercal}$, and $n \times n$ covariance matrix $\bOmega$ having entries $\bOmega_{i,i'}=K(\bx_i,\bx_{i'})$, for every  $i=1, \ldots, n$ and $i'=1, \ldots, n$. Model~\eqref{eq1} has attracted a considerable interest due to its flexibility and its direct connection with binary discrete choice models based on Gaussian latent utilities  $z_i=f(\bx_i)+ \varepsilon_i$, with $ \varepsilon_i \sim \mbox{N}(0,1)$, independently for $i=1, \ldots, n$  \citep{Albert_1993}. In fact, $\pr(y_i=1 \mid f(\bx_i))=\Phi(f(\bx_i))=\pr(z_i>0 \mid f(\bx_i))$. In such settings, a main goal of inference is to evaluate the predictive probabilities of new responses $y_{n+1}$ at a given point $\bx_{n+1}$. Recalling \citet[][Ch.\ 3.3]{ras_2006}, such quantities can be defined as
\begin{align}
\label{eq2}
&\pr(y_{n+1}= 1 \mid  \by)=1-\pr(y_{n+1}= 0\mid \by) \\
&{=}{\int} \Phi(f(\bx_{n+1})) \left[{\int} p(f(\bx_{n+1}), \bof(\bX) \mid  \by) \mbox{d} \bof(\bX) \right] \mbox{d} f(\bx_{n+1}), \nonumber
\end{align}
where $p(f(\bx_{n+1}), \bof(\bX) \mid  \by)$ is the joint posterior density of $(f(\bx_{n+1}),\bof(\bX) )$ induced by model \eqref{eq1}, which does not seem to have an obvious closed form due to the apparent absence  of conjugacy between the probit likelihood and the multivariate Gaussian prior for $(f(\bx_{n+1}),\bof(\bX))$ under \eqref{eq1}. This has motivated extensive research to compute the predictive probabilities in probit models with multivariate Gaussian priors either via  Monte Carlo methods relying on samples from $p(f(\bx_{n+1}), \bof(\bX) \mid  \by)$ \citep{neal1999,Albert_1993,de2005, holmes_2006,choudhuri2007,pakman_2014, Durante2018} or by deriving tractable approximations of $p(f(\bx_{n+1}), \bof(\bX) \mid  \by)$ \citep{kuss2005, chu2005,girolami2006, ras_2006,consonni_2007,nic2008,riihimaki2013} that allow simple evaluation of \eqref{eq2}. Such methods provide state-of-the-art solutions in small-to-moderate dimensional settings, but tend to become inaccurate or computationally impractical in higher dimensions \citep{chopin_2017,Johndrow2018,Durante2018, fasano2019asymptotically}. This issue is inherent to probit GPs where, by definition, the dimension of $\bof(\bX)$ is $n$, or slightly lower when there is overlap in locations, with $n$ being relatively large in most studies.

In this article we aim to cover the above gap by providing novel closed-form expressions for the predictive probabilities in probit  GPs along with improved methods to evaluate the involved quantities in high dimensions. More specifically, in Section \ref{sec.2.1} we first derive a closed-form expression for the marginal likelihood $p(\by)$ under model \eqref{eq1}, and then exploit this result to  show that $\pr(y_{n+1}= 1\mid \by)$ can be expressed as the ratio between cumulative distribution functions of multivariate Gaussians with dimensions $n+1$ and $n$, respectively. To overcome the known issues associated with the evaluation of these two quantities in high dimensions \citep[][]{chop_2011,botev_2017,cao2019,cao2020} we introduce an error-reduction technique for computing ratios of Gaussian cumulative distribution functions that builds on the tile-low-rank method in \citet{cao2020}, and substantially reduces the computational time of state-of-the-art strategies such as minimax tilting methods \citep{botev_2017} and Hamiltonian Monte Carlo samplers (STAN) \citep{hoff_2014}, without affecting accuracy. In Section~\ref{sec.2.2}, we further derive an alternative representation of  $\pr(y_{n+1}= 1\mid \by)$, which relies on functionals of multivariate truncated normals, and we address the intractability of such variables in high dimensions by proposing a  variational approximation based on univariate truncated normals which allows accurate and computationally tractable evaluation of predictive probabilities in high-dimensional contexts. As clarified in Section \ref{sec.2.2}, this solution is computationally more scalable than currently-implemented expectation-propagation approximations \citep[e.g.,][]{kuss2005, chu2005,riihimaki2013}, and improves the accuracy of routinely-used variational solutions \citep[e.g.,][]{girolami2006}, that commonly rely on more restrictive mean-field assumptions, than those required under the proposed approximation. These results are also related to the conditional distribution of the GP given the binary responses, which we show to coincide with a unified skew-normal  (SUN) \citep{arellano_2006} by adapting recent results in \citet{Durante2018}  on Bayesian probit regression. The magnitude of the improvements provided by the new methods presented in Sections \ref{sec.2.1}--\ref{sec.2.2} relative to state-of-the-art competitors is illustrated in simulations in Section \ref{sec.3}, and in an environmental application to Saudi Arabia windspeed in Section \ref{sec.4}. Section \ref{sec.5} contains concluding remarks, whereas all the proofs can be found in the Appendix A. Complete  \texttt{R} code to implement the proposed methods and quantify the improvements relative to state-of-the-art competitors in simulation studies is  available at \url{https://github.com/danieledurante/PredProbitGP}.

\section{Improved Evaluation of Predictive Probabilities in Probit Gaussian Processes}
\label{sec.2}
Sections \ref{sec.2.1} and \ref{sec.2.2} present novel expressions for the predictive probabilities in probit GPs along with improved methods to evaluate the involved quantities efficiently in high dimensions. Feasible grid strategies  to estimate the GP hyperparameters $\balpha$ are also proposed; see Section~\ref{sec.5} for a discussion on the computational tractability of these routines in relation to the dimension of $\balpha$.

\subsection{Evaluation via Gaussian Probability Ratios}
\label{sec.2.1}
To introduce the closed-form expression for $\pr(y_{n+1}= 1\mid \by)$ based on ratios of multivariate Gaussian cumulative distribution functions, first note that by leveraging known properties of Gaussian variables, the probit likelihood in \eqref{eq1} can be written as 
\begin{equation*}
\begin{split}
p(\by \mid \bof(\bX))&= \prod\nolimits_{i=1}^n \Phi(f(\bx_i))^{y_i}[1-\Phi(f(\bx_i)) ]^{1-y_i}\\
&= \prod\nolimits_{i=1}^n \Phi[(2y_i-1)f(\bx_i)]=\Phi_n(\bD\bof(\bX); \bI_n),
\end{split}
\end{equation*}
 where  $\Phi_n(\bD\bof(\bX); \bI_n)$ is the cumulative distribution function of a zero-mean $n$-variate Gaussian with identity covariance matrix $\bI_n$, evaluated at $\bD\bof(\bX)$, with $\bD=\mbox{diag}[(2y_1-1), \ldots, (2y_n-1)]$. Leveraging this form and adapting results in Lemma 7.1 of \citet{azzalini_2013} to our setting, we can easily express the marginal likelihood under model \eqref{eq1} as
\begin{equation}
\begin{split}
p(\by)&={\int} \Phi_n(\bD\bof(\bX); \bI_n) \phi_n(\bof(\bX)-\bxi; \bOmega)\mbox{d} \bof(\bX) \\
&= \Phi_n(\bD\bxi; \bI_n+\bD\bOmega\bD^{\intercal}).
\end{split}
\label{eq3}
\end{equation}
As it will be discussed later on in this article, equation \eqref{eq3} provides a closed-form expression that can be useful to estimate the GP hyperparameters $\balpha$ via direct maximization of $p(\by)$. In addition, as shown in Proposition \ref{prop1}, equation \eqref{eq3}  also allows to derive closed-form expressions for $\pr(y_{n+1}= 1 \mid \by)$.

\begin{Proposition}
Under model \eqref{eq1}, the predictive probability for a new binary response $y_{n+1} \in \{0;1\}$ with predictor $\bx_{n+1} \in \Bbb{R}^q$ is 
\begin{equation}
\begin{split}
{\normalfont \pr}(y_{n+1}= 1 \mid \by)&=1-{\normalfont \pr}(y_{n+1}= 0 \mid \by)\\
&=\frac{\Phi_{n+1}(\bD^*\bxi^*; \bI_{n+1}+\bD^*\bOmega^*\bD^{*\intercal})}{\Phi_n(\bD\bxi; \bI_n+\bD\bOmega\bD^{\intercal})},
\end{split}
\label{eq4}
\end{equation}
with $\bxi^*=[\bxi^{\intercal},m(\bx_{n+1})]^{\intercal}$, $\bD^*=\mbox{\normalfont diag}[(2y_1-1), \ldots, (2y_n-1), 1]$, whereas $\bOmega^*$ is obtained by including one additional row and column to $\bOmega$, which are defined as $\bOmega_{[n+1, \cdot]}^{*\intercal}=\bOmega^*_{[\cdot, n+1]}=[K(\bx_{n+1},\bx_1), \ldots, K(\bx_{n+1},\bx_n), K(\bx_{n+1},\bx_{n+1})]^{\intercal}$.
\label{prop1}
\end{Proposition}

In order to prove Proposition \ref{prop1}, it is sufficient to notice that, by the Bayes rule, $\pr(y_{n+1}= 1 \mid \by)=\pr(y_{n+1}= 1, \by)/p(\by)$ where $\pr(y_{n+1}= 1, \by)$ and $p(\by)$ are the marginal likelihoods of $(y_{n+1}= 1, \by)$ and $\by$, respectively, under model \eqref{eq1}. Replacing such quantities with their closed-form expression as in \eqref{eq3}, leads to \eqref{eq4}. See  Appendix A for a more detailed proof which also includes additional clarifications on equation \eqref{eq3}.

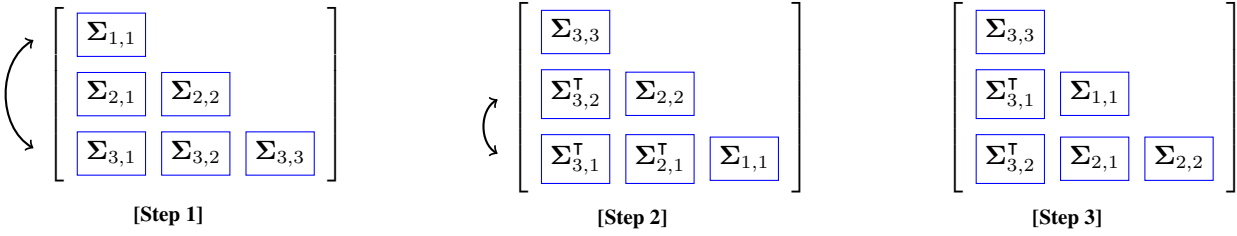
\begin{figure*}[t]
\minipage{0.34\textwidth}
\begin{tikzpicture}[baseline=(U.center)]
\matrix [matrix of math nodes,left delimiter=[,right delimiter={]},nodes={outer sep=1pt},cells={nodes={draw=blue}}] (U) { 
\bSigma_{1,1} &[0.2cm]     &[0.2cm]       \\[0.2cm]
\bSigma_{2,1}  &     \bSigma_{2,2}     &        \\[0.2cm]
\bSigma_{3,1}   &   \bSigma_{3,2}    &      \bSigma_{3,3}      \\
};
\foreach \x/\y/\z in {1/1/1,2/k/j,3/q/t}{
\node[blue,left = 3 mm of U-\x-1] (U-\x-4) {};
\node[red,below= 1 mm of U-3-\x] (U-4-\x) {};
};
\draw[<->, thick, black, bend right=60] (U-1-4) to  node[right, pos=0.5]    {} (U-3-4);
\end{tikzpicture}
\put(-80,-48){\footnotesize \bf [Step 1]}
\endminipage\hfill
\minipage{0.32\textwidth}
\begin{tikzpicture}[baseline=(U.center)]
\matrix [matrix of math nodes,left delimiter=[,right delimiter={]},nodes={outer sep=1pt},cells={nodes={draw=blue}}] (U) { 
\bSigma_{3,3} &[0.2cm]     &[0.2cm]       \\[0.2cm]
\bSigma_{3,2}^{\intercal}  &     \bSigma_{2,2}     &        \\[0.2cm]
\bSigma_{3,1}^{\intercal}    &   \bSigma_{2,1}^{\intercal}    &      \bSigma_{1,1}      \\
};
\foreach \x/\y/\z in {1/1/1,2/k/j,3/q/t}{
\node[blue,left = 3 mm of U-\x-1] (U-\x-4) {};
\node[red,below= 1 mm of U-3-\x] (U-4-\x) {};
};
\draw[<->, thick, black, bend right=60] (U-2-4) to  node[right, pos=0.5]    {} (U-3-4);
\end{tikzpicture}
	\put(-80,-48){\footnotesize \bf [Step 2]}
\endminipage\hfill
\minipage{0.32\textwidth}
\begin{tikzpicture}[baseline=(U.center)]
\matrix [matrix of math nodes,left delimiter=[,right delimiter={]},nodes={outer sep=1pt},cells={nodes={draw=blue}}] (U) { 
\bSigma_{3,3} &[0.2cm]     &[0.2cm]       \\[0.2cm]
\bSigma_{3,1}^{\intercal}  &     \bSigma_{1,1}     &        \\[0.2cm]
\bSigma_{3,2}^{\intercal}    &   \bSigma_{2,1}    &      \bSigma_{2,2}      \\
};
\foreach \x/\y/\z in {1/1/1,2/k/j,3/q/t}{
\node[blue,left = 3 mm of U-\x-1] (U-\x-4) {};
\node[red,below= 1 mm of U-3-\x] (U-4-\x) {};
};
\end{tikzpicture}
	\put(-80,-48){\footnotesize \bf [Step 3]}
\endminipage\hfill
\vspace{-2pt}
	\caption{{\footnotesize{Illustration of the block-reordering strategy  \citep{cao2020}}. {\bf [Step 1]:} Compute $\mbox{min}\{\Phi_{n_1}(\ba_1,\bb_1;\bSigma_{1,1}); \Phi_{n_2}(\ba_2,\bb_2;\bSigma_{2,2}); \Phi_{n_3}(\ba_3,\bb_3;\bSigma_{3,3})\}$, and suppose that the solution is $\Phi_{n_3}(\ba_3,\bb_3;\bSigma_{3,3})$. Then, switch 1st and 3rd block rows and columns, and perform univariate reordering for $\bSigma_{3,1}, \bSigma_{3,2}$ and $\bSigma_{3,3}$.  {\bf [Step 2]:} Compute $\mbox{min}\{\Phi_{n_1}(\ba_1,\bb_1;\bSigma_{1,1}); \Phi_{n_2}(\ba_2,\bb_2;\bSigma_{2,2})\}$, and suppose that the solution is $\Phi_{n_1}(\ba_1,\bb_1;\bSigma_{1,1})$. Then, switch 2nd and 3rd block rows and columns, and perform univariate reordering for $\bSigma_{3,1}, \bSigma_{2,1}$ and $\bSigma_{1,1}$. {\bf [Step 3]:}  Perform univariate reordering for $\bSigma_{2,1}, \bSigma_{3,2}$ and $\bSigma_{2,2}$.}}
	\vspace{-7pt}
	\label{fig:F0}
\end{figure*}

Evaluation of \eqref{eq4} requires the calculation of cumulative distribution functions of multivariate Gaussians, which is known to be a challenging task in high dimensions \citep[][]{genz_1992,chop_2011,botev_2017,genton2018,cao2019,cao2020}. Recent advances via minimax tilting  \citep{botev_2017} allow accurate evaluation of such quantities, but face an increased computational cost which makes such strategies rapidly impractical as $n$ grows. A possible solution to this issue can be found in the separation-of-variable (SOV) algorithm originally introduced by \citet{genz_1992}, and subsequently improved in terms of scalability by \citet{cao2020}. Such a routine decomposes the generic multivariate Gaussian probability $\Phi_n(\ba,\bb;\bSigma)=\int^{\bb}_{\ba} \phi_n(\bu; \bSigma)\mbox{d}\bu$ as
\begin{align}
&\Phi_n(\ba,\bb;\bSigma)=(e_1-d_1) {\int^1_{0}}(e_2-d_2) {\cdots}  {\int^1_{0}}(e_n-d_n)  {\int^1_{0}}\mbox{d} \bw \nonumber \\
&=\mathbb{E}_{\bw}[(e_1-d_1) \cdots (e_n-d_n)]=\mathbb{E}_{\bw}\left[\prod\nolimits_{i=1}^n(e_i-d_i)\right],
\label{eq5}
\end{align}
with $\bw=(w_1, \ldots, w_{n-1})^{\intercal}$ denoting a vector of uniform entries $w_j  \sim \mbox{U}(0,1)$, for $j=1, \ldots, n-1$, whereas
\begin{align*}
d_i &= \Phi\left(\left[a_i - \sum\nolimits_{j=1}^{i-1}l_{ij} \Phi^{-1}[d_j + w_j(e_j - d_j)]\right]l^{-1}_{ii}\right),\\
e_i &= \Phi\left(\left[b_i - \sum\nolimits_{j=1}^{i-1}l_{ij} \Phi^{-1}[d_j + w_j(e_j - d_j)]\right]l^{-1}_{ii}\right),
\end{align*}
for $i=1, \ldots, n$, where  $l_{ij}$ is the $(ij)$-th coefficient in the lower Cholesky factor of $\bSigma$. This decomposition transforms the integration region into the unit hypercube, thus allowing the evaluation of  $\Phi_n(\ba,\bb;\bSigma)$ via functionals of uniform densities.   To further improve the quality of the above estimator, more recent implementations \citep{trinh2015} combine \eqref{eq5} with a univariate reordering  preconditioner that rearranges the integration variables and produces the corresponding Cholesky factor simultaneously at the same $\mathcal{O}(n^3)$ cost of the Cholesky factorization. This prioritization strategy accounts for the width of the integration limits by reordering the variables to ensure that  those having smallest expected values appear as outermost integration variables. Such approach is shown in \cite{trinh2015} and \cite{cao2020} to improve the Monte Carlo convergence rate of \eqref{eq5}, whose integrand is evaluated $R$ times --- corresponding to the Monte Carlo sample size --- each of which has a cost of $\mathcal{O}(n^2)$. Such costs allow the implementation of this strategy in settings with $n \leq 1{,}000$, thus motivating more scalable options in high dimensions. \citet{cao2020} address this issue via a tile-low-rank representation for $\bSigma$ that reduces the cost of the SOV algorithm by substituting the dense matrix-vector multiplication with the low-rank matrix-vector multiplication. A compatible block-reordering is also introduced in place of the univariate reordering to improve the convergence rate at the same cost of the low-rank Cholesky factorization. Specifically, the block-reordering orders the integration variables on the block level based on crude estimates of the block-wise marginal probabilities as shown in Figure~\ref{fig:F0}. Both the block-reordering and the tile-low-rank  version of the SOV algorithm reach their optimal complexities of $\mathcal{O}(n^{5/2})$ and $\mathcal{O}(n^{3/2})$, respectively, when the block size in the tile-low-rank representation is $n^{1/2}$, thus reducing the computational complexity of the classical  SOV algorithm by $n^{1/2}$, and allowing implementation in  tens of thousands of dimensions. 

Although these techniques can be effectively implemented to evaluate multivariate Gaussian probabilities as in \eqref{eq3}, the calculation of ratios among such quantities as in \eqref{eq4} typically requires higher accuracy. Unfortunately, as discussed in \citet{botev_2017} and \citet{cao2020}, the estimation errors of tail multivariate Gaussian probabilities, that also include the cumulative distribution function, can be as large as the probability estimates themselves when $n$ is in hundreds to thousands of dimensions, thus producing unreliable ratio estimates. To address this issue, we propose an error-reduction technique that avoids computing  the numerator and the denominator in \eqref{eq4}  separately, but combines their evaluation under the tile-low-rank representation. Indeed, as is clear from Proposition \ref{prop1}, the denominator in \eqref{eq4} coincides with the numerator without the last integration variable. Hence, keeping the general notation of the SOV algorithm and leveraging \eqref{eq5}, expression \eqref{eq4} can be re-written in the generic form
\begin{equation}
\begin{split}
&\frac{\Phi_{n+1}(\ba,\bb;\bSigma)}{\Phi_{n}(\ba_{-(n+1)},\bb_{-(n+1)};\bSigma_{-(n+1)})}\\ 
&\qquad =\frac{\mathbb{E}_{\bw}[[\prod_{i=1}^n(e_i-d_i)] \cdot(e_{n+1}-d_{n+1})]}{\mathbb{E}_{\bw_{-n}}[\prod_{i=1}^n(e_i-d_i)]},
\end{split}
\label{eq6}
\end{equation}
where $e_i$ and $d_i$ are defined as in equation \eqref{eq5} for $i=1, \ldots, n+1$, whereas $\ba_{-(n+1)}$, $\bb_{-(n+1)}$ and $\bw_{-n}$ are obtained by removing the $(n+1)$-th element in both $\ba$ and $\bb$, and the $n$-th entry in $\bw$, respectively. Similarly, $\bSigma_{-(n+1)}$ coincides with $\bSigma$ without the $(n+1)$-th row and column. As is clear from \eqref{eq6}, the quantities $(e_1-d_1), \ldots, (e_{n}-d_{n})$ are the  same deterministic functions of $\bw$ both in the numerator and in the denominator, and hence, using the same set of Monte Carlo samples $\bw$ in the $n$-dimensional hypercube for estimating the two expectations could significantly reduce the estimation error of their ratio. In particular, our proposed ratio estimator is
\begin{equation}
\begin{split}
&\hat{\pr}(y_{n+1}= 1 \mid \by)\\
&=\frac{\frac{1}{R}\sum_{r=1}^R[[\prod_{i=1}^n(e^{(r)}_i-d^{(r)}_i)] \cdot (e^{(r)}_{n+1}-d^{(r)}_{n+1})]}{\frac{1}{R}\sum_{r=1}^R[\prod_{i=1}^n(e^{(r)}_i-d^{(r)}_i)]},
\end{split}
\label{eq7}
\end{equation}
where  the generic quantities $e^{(r)}_i= e_i(\bw^{(r)})$ and $d^{(r)}_i= d_i(\bw^{(r)})$ denote the values of $e_i$ and $d_i$ in \eqref{eq5} evaluated at the Monte Carlo  sample $\bw^{(r)}$ of $\bw$. Hence, $e^{(r)}_i=e_i(\bw^{(r)}_{-n})$ and $d^{(r)}_i=d_i(\bw^{(r)}_{-n})$ for every  $i=1, \ldots,n$, whereas for  unit $n+1$ these quantities are defined as  $e^{(r)}_{n+1}=e_{n+1}(\bw^{(r)})$ and $d^{(r)}_{n+1}=d_{n+1}(\bw^{(r)})$. Estimator \eqref{eq7} is asymptotically unbiased because the numerator and the denominator converge to $\mathbb{E}_{\bw}[(e_1-d_1)\cdots (e_{n}-d_{n}) \cdot (e_{n+1}-d_{n+1})]$ and $\mathbb{E}_{\bw_{-n}}[(e_1-d_1)\cdots (e_{n}-d_{n})]$, respectively, and hence equation \eqref{eq7} converges to \eqref{eq6} in probability. Moreover, equation~\eqref{eq7} is guaranteed to be in $(0, 1)$, thus producing an estimator whose variance is always smaller than $0.25$. This is not the case when  the numerator and the denominator in \eqref{eq4} are estimated separately. Indeed, as discussed in \citet{botev_2017} and  \citet{cao2020}, when $n$ is high the estimation errors of the two cumulative distribution functions in \eqref{eq4} are often as large as the estimates themselves, thus producing estimated ratios possibly outside of the range $(0, 1)$, and with high variance.

\begin{algorithm*}[t]
	\caption{Compute \eqref{eq4} via the estimator \eqref{eq7}} 
{\small	\vspace{3pt}
{\bf [a]} Set $\ba=-\infty$, $\bb=\bD^*\bxi^*$, $\bSigma= \bI_{n+1}+\bD^*\bOmega^*\bD^{*\intercal}$, and draw $\bw^{(1)}, \ldots, \bw^{(R)}$ uniform samples from the unit hypercurbe in $(0,1)^n$.\\
\vspace{3pt}
		{\bf [b]} Apply  block-reordering \citep{cao2020} to ($\ba_{-(n+1)}$, $\bb_{-(n+1)}$, $\bSigma_{-(n+1)}$), which produces the tile-low-rank Cholesky factor $\bL_{-(n+1)}$, and the reordered $\ba_{-(n+1)}$ and $\bb_{-(n+1)}$.\\
\vspace{3pt}
		{\bf [c]} Compute $\bL_{[n+1,1:n+1]}$ using $\bSigma$ and $\bL_{-(n+1)}$.\\		
\vspace{3pt}	
{\bf [d]} Obtain the quantities required to evaluate equation \eqref{eq7}. \\		
\vspace{2pt}	
		\For{$r=1,\ldots,R$}{
			\vspace{5pt}		
		{\bf [d.1]} Compute $e_i(\bw^{(r)}_{-n})-d_i(\bw^{(r)}_{-n})$, for every statistical unit  $i=1, \ldots, n,$ by applying the tile-low-rank variant of~\eqref{eq5} \citep{cao2020} to ($\ba_{-(n+1)}$, $\bb_{-(n+1)}$, $\bSigma_{-(n+1)}$). \\ Store also the vector $\bv^{(r)}=[\Phi^{-1}(d_1(\bw^{(r)}_{-n})+w^{(r)}_1[e_1(\bw^{(r)}_{-n})-d_1(\bw^{(r)}_{-n})]), \ldots, \Phi^{-1}(d_n(\bw^{(r)}_{-n})+w^{(r)}_n[e_n(\bw^{(r)}_{-n})-d_n(\bw^{(r)}_{-n})])]^{\intercal}$. \\
	\vspace{3pt}		
	{\bf [d.2]} Set $e_{n+1}(\bw^{(r)})-d_{n+1}(\bw^{(r)})=\Phi(\frac{b_{n+1}- \bL_{[n+1,1:n]}\bv^{(r)}}{l_{n+1,n+1}})-\Phi(\frac{a_{n+1}- \bL_{[n+1,1:n]}\bv^{(r)}}{l_{n+1,n+1}})$.
	}
		\vspace{3pt}		
	{\bf [e]}  Estimate~\eqref{eq4} via Monte Carlo as in equation~\eqref{eq7} using the quantities computed in step {\bf [d]}.
			\vspace{2pt}	}	
		\label{algo:1}
\end{algorithm*}

The pseudo-code to evaluate~\eqref{eq4} via the estimator presented in~\eqref{eq7} is provided in Algorithm~\ref{algo:1}. In step~{\bf[b]} of Algorithm~\ref{algo:1}, the block-reordering produces a new variable order which is used to reorder the integration limits, whereas in step~{\bf[c]} the inverse matrices of the diagonal blocks of the tile-low-rank Cholesky factor computed in step~{\bf[b]} are recycled to maximize efficiency. Also the quantities in~{\bf[d.1]} do not need to be re-evaluated every time a new prediction is required since they only depend on the observed training data, and hence such quantities can be pre-computed and stored separately. This yields an overall computational complexity of $\mathcal{O}(n^{5/2}+Rn^{3/2})$ for Algorithm~\ref{algo:1}, which comprises the $\mathcal{O}(n^{5/2})$ pre-computation cost of the block-reordering strategy to produce the tile-low-rank Cholesky factor, and the $\mathcal{O}(n^{3/2})$ operations per sample to compute the quantities in step {\bf[d]}. This allows to reduce the overall complexity of other state-of-the-art accurate alternatives for evaluating \eqref{eq4}, such as the strategy proposed by \citet{botev_2017}, that has an $\mathcal{O}(n^{3})$ pre-computation cost for obtaining the minimax exponentially-tilted estimate, and then requires $\mathcal{O}(n^{2})$ matrix-vector multiplication operations per sample, for a total of $\mathcal{O}(n^{3}+Rn^{2})$.

The computational gains achieved under Algorithm~\ref{algo:1} are also inherited when adapting the method in \citet{cao2020} to evaluate the marginal likelihood in equation \eqref{eq3}, thereby facilitating the development of feasible estimation strategies for the GP hyperparameters $\balpha$ via the maximization of $p(\by)$. Although this task is amenable to a variety of gradient-based optimization algorithms, in practice, the implementation of these routines, might be subject to computational bottlenecks and tedious calculations which involve derivatives of multivariate Gaussian cumulative distribution functions. To circumvent these issues, we propose to rely on a heuristic grid search strategy which evaluates $p(\by)$ at several reasonable combinations of $\balpha$ values, and then selects as estimate the configuration yielding the highest marginal likelihood. As highlighted in Section~\ref{sec.1}, $\balpha$ comprises few hyperparameters in routine GP implementations, and prediction is typically robust to minor variations in $\balpha$, thereby making these grid strategies practically feasible and still reliable in common applications \citep[e.g.,][]{kuss2005,ras_2006,nic2008,riihimaki2013}; see also the final discussion in Section~\ref{sec.5}  for additional details and possible solutions regarding the computational bottlenecks of the proposed grid search in situations when the number of hyperparameters in $\balpha$ is moderate-to-large.

\subsection{Evaluation via Functionals of  Truncated Normals}
\label{sec.2.2}
The methodologies  in Section \ref{sec.2.1} allow substantial improvements in terms of accuracy and scalability in the evaluation of predictive probabilities, but still require to deal with multivariate Gaussian cumulative distribution functions, a challenging task, especially in high dimensions. To overcome this issue, we derive an alternative expression for $\pr(y_{n+1}= 1 \mid \by)$ relying on functionals of multivariate truncated normals which are then approximated via mean-field variational Bayes \citep[e.g.,][]{blei2017} to facilitate simple Monte Carlo evaluation of   $\pr(y_{n+1}= 1 \mid \by)$ using samples from univariate truncated normals. 

To derive this alternative expression, we shall first notice that the joint posterior $p(f(\bx_{n+1}), \bof(\bX) \mid  \by)$ in \eqref{eq2} can be factorized as $p(f(\bx_{n+1}) \mid \bof(\bX)) p(\bof(\bX) \mid \by)$, provided that $f(\bx_{n+1})$ does not appear in the likelihood for $\by$, which is true because there is no overlap among predictors. Exploiting the well-known properties of GPs \citep[][]{ras_2006}, the first factor  in the above expression can be easily derived by applying the closure under conditioning property of multivariate Gaussians, thus obtaining the univariate normal density
\begin{equation}
\begin{split}
& p(f(\bx_{n+1}) \mid \bof(\bX)) \\
&=\phi(f(\bx_{n+1}) -(\mu_{x_{n+1}}+\bH_{x_{n+1}}\bof(\bX)); \sigma_{x_{n+1}}^2 ),
\end{split}
\label{eq8}
\end{equation}
with $\mu_{x_{n+1}}{=} \ m(\bx_{n+1}){-}\bH_{x_{n+1}}\bxi$, $\bH_{x_{n+1}}{=} \ \bOmega^*_{[n+1,1:n]}\bOmega^{-1}$ and $ \sigma_{x_{n+1}}^2= K(\bx_{n+1},\bx_{n+1})-\bOmega^*_{[n+1,1:n]}\bOmega^{-1}\bOmega^*_{[1:n,n+1]} $, where the different quantities entering these expressions are defined as in  \eqref{eq1} and~\eqref{eq4}. By adapting the recent conjugacy results for probit models with Gaussian priors in  \citet{Durante2018} to this GP setting, it is also possible to show that $p(\bof(\bX) \mid \by)$ is the density of the unified skew-normal (SUN) \citep{arellano_2006} $\mbox{SUN}_{n,n}(\bxi,\bOmega,\bar{\bOmega}\bomega\bD^{\intercal}\bs^{-1},\bs^{-1}\bD\bxi, \bs^{-1}(\bD\bOmega\bD^{\intercal}{+}\bI_n)\bs^{-1})$, with $\bs=[(\bD\bOmega\bD^{\intercal}{+}\bI_n)\odot \bI_n]^{1/2}$, $\bar{\bOmega}=\bomega^{-1}\bOmega\bomega^{-1}$ and $\bomega=(\bOmega \odot \bI_n)^{1/2}$. Indeed, recalling the results  in Sections \ref{sec.1}--\ref{sec.2.1} and applying the Bayes rule, we have that $p(\bof(\bX) \mid \by) \propto p(\bof(\bX))p(\by \mid \bof(\bX))=\phi_n(\bof(\bX)-\bxi; \bOmega) \Phi_n(\bD\bof(\bX); \bI_n)$, which is the kernel of a SUN density --- as shown in the proof of Theorem 1 by  \citet{Durante2018}. This class of random variables introduces asymmetric shapes in Gaussian densities via a skewness-inducing mechanism driven by the cumulative distribution function of an $n$-variate Gaussian with a full-rank covariance matrix. Hence, the evaluation of $ p(\bof(\bX) \mid \by)$ still requires calculation of multivariate Gaussian probabilities, leading to the same issues discussed in Section \ref{sec.2.1}; see \citet{arellano_2006}, \citet{azzalini_2013} and \citet{Durante2018} for an in-depth discussion on the properties of SUN variables for posterior inference.

A possibile option to address the above issue is to consider the discrete-choice interpretation of the probit GP introduced in Section \ref{sec.1}. Under this  representation, model \eqref{eq1} can be  re-expressed as $y_i=1(z_i>0)$, with $(z_i \mid f(\bx_i)) \sim \mbox{N}(f(\bx_i),1)$, independently for  $i=1, \ldots, n$, and $\bof(\bX)=(f(\bx_1), \ldots,f(\bx_n))^{\intercal} \sim \mbox{N}_n(\bxi, \bOmega)$. Adapting the results in \citet{holmes_2006} to our GP setting, the joint posterior $p(\bof(\bX), \bz \mid \by)$ of $\bof(\bX)$ and the augmented data $\bz=(z_1, \ldots, z_n)^{\intercal}$, factorizes as $p(\bof(\bX) \mid  \bz) p(\bz \mid \by)$, with
\begin{eqnarray}
\begin{split}
&p(\bof(\bX) \mid \bz) \\
&=\phi_n(\bof(\bX)-(\bOmega^{-1}+\bI_n)^{-1}(\bOmega^{-1}\bxi+\bz); (\bOmega^{-1}+\bI_n)^{-1})\\
&\qquad  =\phi_n(\bof(\bX)-(\bmu_{\bX}+\bSigma_{\bX}\bz); \bSigma_{\bX}), \\
&p(\bz \mid \by)\\
& \propto \phi_n(\bz - \bxi; \bI_n+\bOmega){\textstyle\prod}_{i=1}^n1[(2y_i-1)z_i>0]\\
& \qquad =\phi_n(\bz - \bxi; \bSigma_{\bz}){\textstyle\prod}_{i=1}^n1[(2y_i-1)z_i>0],
\end{split}
\label{eq9}
\end{eqnarray}
where $\bSigma_{\bX}=(\bOmega^{-1}+\bI_n)^{-1}$, $\bmu_{\bX}=\bSigma_{\bX}\bOmega^{-1}\bxi$ and $\bSigma_{\bz}=\bI_n+\bOmega$. Therefore, the joint posterior  density $p(\bof(\bX) \mid  \bz) p(\bz \mid \by)$ factorizes as the product of a Gaussian for $p(\bof(\bX) \mid \bz)$ and a multivariate truncated normal for $p(\bz \mid \by)$ obtained via component-wise truncation of $\mbox{N}_n(\bxi,\bSigma_{\bz})$ below or above $0$, depending on whether $y_i=1$ or $y_i=0$, respectively, for $i=1, \ldots, n$. As shown in Proposition \ref{prop2}, by combining equations \eqref{eq8}--\eqref{eq9} with Lemma 7.1 in \citet{azzalini_2013}, it is possible to obtain an alternative expression for $\pr(y_{n+1}= 1 \mid \by)$ based on functionals of multivariate truncated normals. See the Appendix A for a detailed proof.
\begin{Proposition}
Under model \eqref{eq1}, the predictive probability for a new response $y_{n+1} \in \{0;1\}$ with predictor $\bx_{n+1} \in \Bbb{R}^q$ is 
\begin{equation}
\begin{split}
&{\normalfont \pr}(y_{n+1}= 1 \mid \by)=1-{\normalfont \pr}(y_{n+1}=0\mid \by)\\
&= \mathbb{E}_{\bz \mid \by}[\mathbb{E}_{\bof(\bX) \mid \bz}(\mathbb{E}_{f(\bx_{n+1}) \mid \bof(\bX)}[\Phi(f(\bx_{n+1}))])]\\
&= \mathbb{E}_{\bz \mid \by}[\mathbb{E}_{\bof(\bX) \mid \bz}[\Phi(\mu_{x_{n+1}}+\bH_{x_{n+1}}\bof(\bX);1+\sigma_{x_{n+1}}^2 )]]\\
&= \mathbb{E}_{\bz \mid \by}[\Phi(\mu_{x_{n+1}}+ \bH_{x_{n+1}} (\bmu_{\bX}+ \bSigma_{\bX}\bz);\\
& \qquad \qquad  \qquad \qquad  \quad 1+\sigma_{x_{n+1}}^2+ \bH_{x_{n+1}} \bSigma_{\bX}\bH^{\intercal}_{x_{n+1}})], 
\label{eq10}
\end{split}
\end{equation}
where the quantities in \eqref{eq10} are defined as in equations \eqref{eq8} and \eqref{eq9}, whereas $\mathbb{E}_{\bz \mid \by}(\cdot)$ denotes the expectation with respect to the multivariate truncated normal density $p(\bz \mid \by)$ in \eqref{eq9}.
\label{prop2}
\end{Proposition}

Leveraging Proposition \ref{prop2} it is possible to evaluate $\pr(y_{n+1}= 1 \mid \by)$ via Monte Carlo methods based on independent samples from the multivariate truncated normal with density as in \eqref{eq9}, thus producing the estimate $\hat{\pr}(y_{n+1}= 1 \mid \by)=\sum_{r=1}^R\Phi(\mu_{x_{n+1}}+ \bH_{x_{n+1}} (\bmu_{\bX}+ \bSigma_{\bX}\bz^{(r)});1+\sigma_{x_{n+1}}^2+ \bH_{x_{n+1}} \bSigma_{\bX}\bH^{\intercal}_{x_{n+1}})/R$, where $\bz^{(1)}, \ldots, \bz^{(R)}$ are independent and identically distributed samples from $p(\bz \mid \by)$ in \eqref{eq9}. Unfortunately, sampling from multivariate truncated normals in settings where $n$ is larger than a few hundreds raises the same computational issues discussed in Section \ref{sec.2.1}, i.e., the evaluation of multivariate Gaussian cumulative distribution functions \citep{holmes_2006,botev_2017, pakman_2014,Durante2018,fasano2019asymptotically}. 

\begin{algorithm*}[t]
	\caption{Compute \eqref{eq10} via Monte Carlo  as in \eqref{eq13} based on the mean-field approximation of $p(\bz \mid \by)$} 
	{\small
\vspace{5pt}	
{\bf CAVI algorithm}\\
\vspace{3pt}	
{\bf [a]}  Pre-compute $\bOmega^{-1}$ and $\bSigma^{-1}_{\bz}=(\bI_n+\bOmega)^{-1}$, and leverage the standard properties for the inverse of block matrices to obtain $\bH_{z_i}$ and $ \sigma_{z_i}^{2}$, for each $i=1, \ldots, n$ as suitable sub-blocks of $\bSigma^{-1}_{\bz}$.	\\
\vspace{3pt}
{\bf [b]} Initialize $\bz^{(0)}\in \Bbb{R}^n$, and apply  CAVI  to obtain the optimal mean-field approximation $q^*(\bz)=\prod_{i=1}^n q^*(z_i)$ for $p(\bz \mid \by)$.	\\
\vspace{2pt}
		\For{$t=1$ until convergence}{
		\For{$i=1, \ldots, n$}{
		Set the approximating density for $z_i$ at step $t$ equal to
		$q^{(t)}(z_i) \propto \phi(z_i-[\xi_i+\bH_{z_i}(\bz^{(t-1)}_{-i}-\bxi_{-i})]; \sigma_{z_i}^{2})1[(2y_i-1)z_i>0]$
		 with $\bz^{(t-1)}_{-i}=[\mathbb{E}_{q^{(t)}(z_1)}(z_1), \ldots, \mathbb{E}_{q^{(t)}(z_{i-1})}(z_{i-1}),\mathbb{E}_{q^{(t-1)}(z_{i+1})}(z_{i+1}), \ldots, \mathbb{E}_{q^{(t-1)}(z_{n})}(z_{n})]^{\intercal}$.
		}
		\vspace{2pt}
		{\bf Output:} $q^*(\bz)=\prod_{i=1}^n q^*(z_i)$, where each $q^*(z_i)$ is a univariate truncated normal.
	}
		\vspace{8pt}	
		{\bf Evaluation of predictive probabilities}\\
\vspace{3pt}	
{\bf [c]}  Compute $\bOmega^{-1}\bSigma_{\bX}$ which enters the definition of the key quantities in  \eqref{eq13}, namely $\bH_{x_{n+1}}\bmu_{\bX}$ and $\bH_{x_{n+1}}\bSigma_{\bX}$. Note that, by  standard properties of matrix inverse $\bOmega^{-1}\bSigma_{\bX}=\bOmega^{-1}(\bOmega^{-1}+\bI_n)^{-1}=(\bI_n+\bOmega)^{-1}$, which coincides with $\bSigma^{-1}_{\bz}$ already pre-computed in {\bf [a]}.	\\
\vspace{3pt}	
	{\bf [d]} Estimate  \eqref{eq10} via Monte Carlo as in \eqref{eq13}, based on $R$ independent samples from the optimal univariate truncated normal approximating densities provided by step  {\bf [b]}.
			\vspace{2pt}	}	
		\label{algo:2}
\end{algorithm*}

To avoid these issues, we adapt ideas in \citet{fasano2019asymptotically} and propose to replace the intractable sampling density $p(\bz \mid \by)$ with a mean-field approximation $q^*(\bz)=\prod_{i=1}^n q^*(z_i)$ factorizing over marginals $q^*(z_1), \ldots, q^*(z_n)$. In this way, the Monte Carlo estimate for $\pr(y_{n+1}= 1 \mid \by)$ can be obtained by sampling $R$ times from $n$ independent univariate approximate densities $q^*(z_1), \ldots, q^*(z_n)$ instead of the exact but intractable joint density $p(\bz \mid \by)$. Recalling the classical mean-field variational Bayes (VB) framework \citep[e.g.,][]{blei2017}, the optimal approximating density $q^*(\bz)$ is the one that minimizes the Kullback--Leibler (KL) divergence $\textsc{kl}[q(\bz) \| p(\bz \mid \by)]=\mathbb{E}_{q(\bz)}\{\log [q(\bz)/p(\bz \mid \by)]\}$ \citep{kullback1951} to $p(\bz \mid \by)$ among all the densities within the mean-field family $\mathcal{Q}=\{q(\bz): q(\bz)=\prod_{i=1}^n q(z_i) \}$. The solution of such a minimization problem is, typically, not available in closed form but can be obtained via coordinate ascent variational inference (CAVI) algorithms \citep[][]{bishop2006,blei2017} that iteratively minimize the KL with respect to each component $q(z_i)$ at a time, keeping fixed  the others at their most recent estimate $\bq^{(t-1)}(\bz_{-i})$, where $\bz_{-i}$ denotes vector $\bz$ without the $i$-th entry. Recalling \citet{bishop2006}, this is accomplished via the updates
\begin{equation}
q^{(t)}(z_i) \propto \exp[ \mathbb{E}_{\bq^{(t-1)}(\bz_{-i})}(\log[p(z_i \mid \bz_{-i},\by)])], 
\label{eq11}
\end{equation}
for each  $i=1, \ldots,n$, at iteration $t$, until convergence. In \eqref{eq11}, the quantity  $p(z_i \mid \bz_{-i},\by)$ denotes the full conditional density of $z_i$. Due to the closure under conditioning property of the multivariate truncated normal \citep{horrace2005}, such a quantity can be derived explicitly from $p(\bz \mid \by)$ in \eqref{eq9} and coincides with the density of a univariate truncated normal. In particular, we can express each $p(z_i \mid \bz_{-i},\by)$ as
\begin{align}
&p(z_i \mid \bz_{-i},\by) \label{eq12} \\
&\propto\phi(z_i-[\xi_i+\bH_{z_i}(\bz_{-i}-\bxi_{-i})]; \sigma_{z_i}^{2})1[(2y_i-1)z_i>0],
\nonumber
\end{align}
where $\bxi_{-i}$ denotes the prior mean vector $\bxi$ without the $i$-th element, whereas $\bH_{z_i}=\bSigma_{\bz[i,-i]}(\bSigma_{\bz[-i,-i]})^{-1}$ and $ \sigma_{z_i}^{2}=\bSigma_{\bz[i,i]}-\bSigma_{\bz[i,-i]}(\bSigma_{\bz[-i,-i]})^{-1}\bSigma_{\bz[-i,i]}$. Density in \eqref{eq12} has a log-kernel which is linear in $\bz_{-i}$ and, therefore, replacing the expression for $p(z_i \mid \bz_{-i},\by)$ within the CAVI updates in equation \eqref{eq11}, it follows that also $q^{(t)}(z_i)$ has a univariate truncated normal density as in  \eqref{eq12} with $\bz_{-i}$ replaced by 
\begin{equation*}
\begin{split}
\bz_{-i}^{(t-1)}=[&\mathbb{E}_{q^{(t)}(z_1)}(z_1), \ldots, \mathbb{E}_{q^{(t)}(z_{i-1})}(z_{i-1}),\\
&\mathbb{E}_{q^{(t-1)}(z_{i+1})}(z_{i+1}), \ldots, \mathbb{E}_{q^{(t-1)}(z_{n})}(z_{n})]^{\intercal}.
\end{split}
\end{equation*}
Each term in $\bz^{(t-1)}_{-i}$ is the expectation of a univariate   truncated normal, that is explicitly available, thus producing a simple CAVI relying on closed-form updates; see Algorithm~\ref{algo:2}.

Once the optimal univariate truncated normal approximating densities $q^*(z_1), \ldots, q^*(z_n)$ are available, equation \eqref{eq10} can be easily evaluated via Monte Carlo by letting  
\begin{equation}
\begin{split}
		&\hat{\pr}(y_{n+1}= 1 \mid \by)\\
		&\qquad =\frac{1}{R}\sum\nolimits_{r=1}^R\Phi(\mu_{x_{n+1}}{+} \ \bH_{x_{n+1}} (\bmu_{\bX}+ \bSigma_{\bX}\bz^{*(r)});\\
		&\   \qquad  \qquad  \qquad  \qquad  \qquad 1+\sigma_{x_{n+1}}^2{+} \ \bH_{x_{n+1}} \bSigma_{\bX}\bH^{\intercal}_{x_{n+1}}),
		\end{split}
		\label{eq13}
		\end{equation}
		with $\bz^{*(r)}=(z_1^{*(r)}, \ldots, z_n^{*(r)})^{\intercal}$, where each $z_i^{*(r)}$ can be efficiently sampled from the corresponding univariate truncated normal approximating density $q^*(z_i)$, independently for $i=1, \ldots, n$ and $r=1, \ldots, R$. Unlike for the multivariate case, sampling from univariate truncated normals can be effectively done in standard statistical softwares, thus avoiding issues in large $n$ settings. 
		
Algorithm~\ref{algo:2} provides the pseudo-code to implement the proposed VB approximation for the predictive probabilities in \eqref{eq10}. As is clear from  Algorithm~\ref{algo:2}, the quantities 	$\bH_{z_i}$ and $ \sigma_{z_i}^{2}$, $i=1, \ldots, n$, involved in step {\bf [b]}, coincide with suitable sub-blocks of $\bSigma^{-1}_{\bz}$. Due to this, the operations required to update each $q^{(t)}(z_i)$ in {\bf [b]} are linear in $n$, and, therefore, the overall cost of CAVI is $\mathcal{O}(n^3)$, which coincides with the cost for pre-computing matrix $\bSigma^{-1}_{\bz}$ in {\bf [a]}. Leveraging these results, the evaluation of the predictive probabilities in step  {\bf [d]} implies an  $\mathcal{O}(n)$ cost per Monte Carlo sample, since, according to step {\bf [c]}, the main quantities  in \eqref{eq13} can be derived from those pre-computed in {\bf [a]}. This yields a total cost for Algorithm~\ref{algo:2}  of $\mathcal{O}(n^3+Rn)$ which reduces by $n^{1/2}$ the Monte Carlo complexity of Algorithm~\ref{algo:1}, but increases by the same amount the pre-computation cost. As for Algorithm~\ref{algo:1}, also in Algorithm~\ref{algo:2} the most computationally intensive steps in {\bf [a]}--{\bf [c]} do not need to be re-executed each time a new prediction is required, thereby making computation of predictive probabilities at multiple data points almost as expensive as implementing this task for a single location. 
		
As discussed in e.g., \citet{kuss2005}, \citet{riihimaki2013}, the cubic cost is commonly unavoidable in standard GP settings with generic covariance matrix. However, unlike for alternative approximations relying, for instance, on expectation-propagation (EP) methods \citep[e.g.,][]{kuss2005,riihimaki2013}, this $\mathcal{O}(n^3)$ cost is only paid once in the pre-computation step, and not for each iteration of the optimization routine. This yields substantial improvements in terms of scalability to high dimensions relative to EP. As outlined in the simulation studies in Section \ref{sec.3}, these gains are obtained without sacrificing estimation accuracy, when compared to non-approximate methods. This is due to the fact that the proposed  strategy integrates out $\bof(\bX)$ analytically in \eqref{eq10} with respect to its exact density $p(\bof(\bX) \mid \bz)$, and only approximates $p(\bz \mid \by)$. This departs from classical VB solutions \citep{girolami2006} which consider a mean-field approximation $q^*(\bof(\bX))\prod_{i=1}^nq^*(z_i)$ of the joint  density $p(\bof(\bX), \bz \mid \by)$, and then compute predictive probabilities based on Monte Carlo samples from $q^*(\bof(\bX))$. This yields  less accurate estimates of the predictive probabilities that, unlike for the solution we propose, do not  fully incorporate the exact dependence between $\bof(\bX)$ and $\bz$ (\citealp[e.g.,][Figure 6]{nic2008};  \citealp{fasano2019asymptotically}).

\section{Simulation Studies}
\label{sec.3}

\begin{figure*}[t]
	\centering
		\includegraphics[scale=.65]{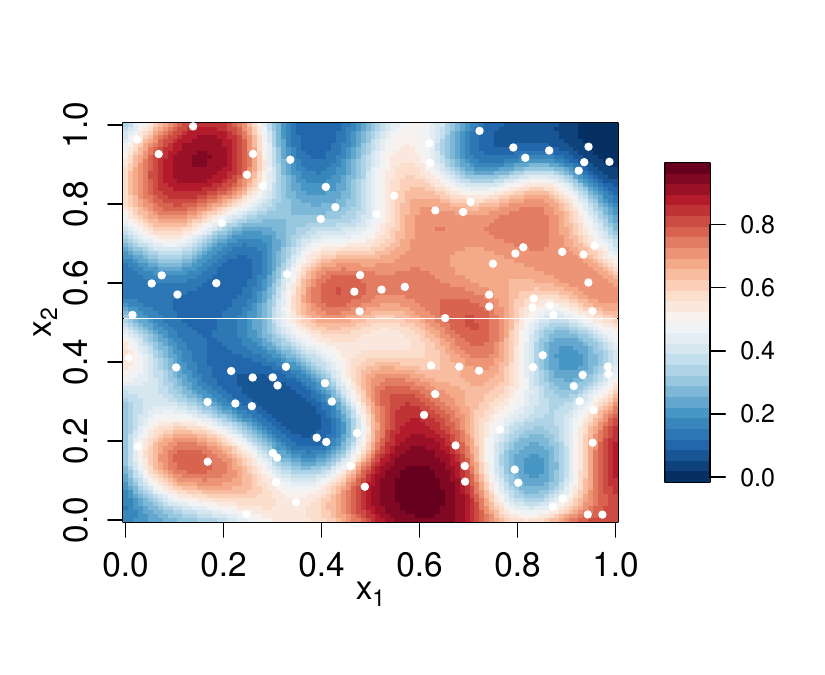}
		\includegraphics[scale=.65]{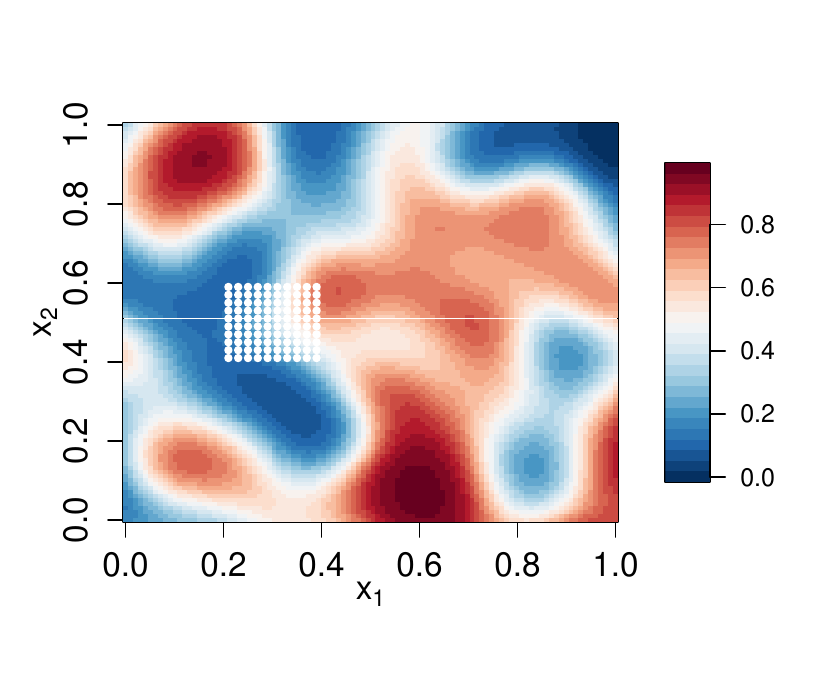}	
		\vspace{-18pt}	
	\caption{{\footnotesize{Simulated probabilities on the $100 \times 100$ grid $\mathcal{G}=\{\bx=(x_1{,}x_2): x_1 \in (1/100,2/100, \ldots, 100/100),  x_2 \in (1/100,2/100, \ldots, 100/100)\}$ in the unit square, where $f(\bx)$ is a zero mean GP with squared exponential covariance kernel. White circles denote the 100 test locations distributed randomly (left) and on a grid (right), used for prediction.}}}
	\label{fig:F1}
\vspace{-5pt}
\end{figure*}

In this section, we study the gains in accuracy and computational scalability of the methods developed in Sections \ref{sec.2.1} and \ref{sec.2.2} relative to state-of-the-art alternatives. More specifically, to quantify the magnitude of the improvements provided by the tile-low-rank (TLR) strategy developed in Section \ref{sec.2.1}, we consider as a competitor the recent minimax tilting  method (TN) by \citet{botev_2017} (see \texttt{R} package \texttt{TruncatedNormal}), which is used here to evaluate the Gaussian cumulative distribution functions involved in the predictive probability \eqref{eq4}. This strategy has been shown to substantially improve the accuracy and computational tractability of other state-of-the-art solutions and, hence, provides a challenging benchmark to assess the gains of the TLR procedure. The performance improvements of the VB  developed in Section~\ref{sec.2.2} are, instead, compared against Monte Carlo inference under the widely-used STAN implementation of the Hamiltonian no-u-turn sampler \citep{hoff_2014} available in the state-of-the-art \texttt{R} package \texttt{rstan}. Both VB and  STAN provide Monte Carlo estimates of predictive probabilities but, unlike for our proposed  VB solution, STAN relies on samples from the exact posterior, thus providing a relevant and routinely-used competitor for evaluating the accuracy of the proposed VB approximation and its gains in runtime. As discussed in Section~\ref{sec.2.2}, classical mean-field variational methods \citep[e.g.,][]{girolami2006} and EP solutions \citep[e.g.,][]{kuss2005,riihimaki2013} would  yield reduced accuracy or higher computational costs than the proposed VB, and, hence, are not implemented. 

To evaluate the performance in high dimensional settings, we generate the binary response data on the $100 \times 100$ unit grid $\mathcal{G}=\{\bx=(x_1,x_2): x_1 \in (1/100,2/100, \ldots, 100/100),  x_2 \in (1/100,2/100, \ldots, 100/100)\}$ with equally-spaced predictors, thereby obtaining $n=10{,}000$ non-overlapping configurations. At these locations, we simulate $y_1, \ldots, y_{10{,}000}$ from independent Bernoullis with probabilities $\Phi(f_0(\bx_1)), \ldots, \Phi(f_0(\bx_{10{,}000}))$ displayed in Figure \ref{fig:F1}, where $\bof_0(\bX)=(f_0(\bx_1), \ldots, f_0(\bx_{10{,}000}))^{\intercal}$ is a sample from a GP having mean $m(\bx)=0$ and squared exponential covariance kernel 
\begin{equation*}
K(\bx,\bx')=\exp\{-[\alpha^2_1(x_1-x'_1)^2 +\alpha^2_2(x_2-x'_2)^2]\},
\end{equation*}
with $\balpha=(\alpha_1,\alpha_2)=(\sqrt{30}, \sqrt{30})$ to illustrate also performance in estimating more than one GP hyperparameter; see also Section~\ref{sec.5} for a discussion on hyperparameter estimation in higher dimensional settings. The proportion of `1's and `0's in the  $10{,}000$ simulated binary responses is $49.5\%$ and $50.5\%$, respectively, thus providing a balanced dataset. To assess performance in estimating the predictive probabilities, we adopt a validation-set approach by simulating probability parameters and the associated binary responses for $100$ out-of-sample units under two scenarios. As outlined in Figure~\ref{fig:F1}, the first one relies on randomly distributed locations, whereas the second focuses on a grid structure, and both comprise relatively balanced binary responses, as for the training sample.  To provide a more comprehensive assessment, we also compare performance in lower-dimensional training problems with $n \in \{15^2; 25^2; 50^2\}$ obtained by selecting a $n^{1/2} \times n^{1/2}$ sub-grid of $\mathcal{G}$ with equally-spaced configurations between $0$ and $1$, along with their associated probability parameters and simulated responses. 

\begin{table*}[t]
\caption{{\footnotesize{Runtimes and accuracy in estimating out-of-sample predictive probabilities, at varying training sample size $n$, of STAN \citep{hoff_2014}, TN \citep{botev_2017}, TLR (Section \ref{sec.2.1}) and VB (Section \ref{sec.2.2}), when the $100$ test locations are distributed either randomly [random] or on a grid [grid]. TIME: runtime in seconds for predicting at one location. MSE: mean squared error between the $100$ estimated predictive probabilities and the true ones. Empty cells refer to situations in which the overall runtime of the whole prediction task exceeded the conservative budget of one day.}}}\label{tab:T1}
\addtolength{\tabcolsep}{-2pt}
	\centering
	\resizebox{10.1cm}{!}{
\begin{tabular}{ll|cccc}
Method& Performance measures&$n={225}$  & $n=625$ & $n=2{,}500$& $n=10{,}000$ \\
\midrule
	STAN		&TIME {\footnotesize{[seconds]}} & 1{,}382 & 18{,}066 & ---& ---\\
		&MSE {\footnotesize{[random]}} & 0.015 & 0.014 & ---& --- \\
		&MSE {\footnotesize{[grid]}} & 0.023 & 0.015 & ---& --- \\
\midrule
	TN		&TIME  {\footnotesize{[seconds]}} & 7 & 41 & --- & ---  \\
		&MSE  {\footnotesize{[random]}} & 0.017 & 0.014 & --- & ---  \\
	&MSE  {\footnotesize{[grid]}} & 0.027 & 0.017 & --- & ---  \\
\midrule
	TLR		&TIME  {\footnotesize{[seconds]}} & 1 & 5 & 37 & 250 \\
		&MSE  {\footnotesize{[random]}} & 0.017 & 0.014 & 0.005& 0.002 \\
	&MSE  {\footnotesize{[grid]}} & 0.025 & 0.019 & 0.007& 0.003\\
\midrule
	VB	&TIME  {\footnotesize{[seconds]}} & 1 & 3 & 23& 898 \\
		&MSE  {\footnotesize{[random]}} & 0.016 & 0.014 &  0.005& 0.001 \\
	&MSE  {\footnotesize{[grid]}} & 0.025 & 0.017 & 0.004& 0.001 \\
\bottomrule
\end{tabular}}
\end{table*}

Table \ref{tab:T1} summarizes the accuracy and computational scalability of the methods analyzed, at varying $n$ and under the two different scenarios considered for prediction. In reporting the results, we set conservative computational budget of one day and  compute the out-of-sample validation MSEs instead of the cross-validated ones to limit the overall computational effort within our capacity, especially for the two competitors TN and  STAN. To provide a reliable comparison between the different implementations, we consider the runtime for predicting one test unit. Such a measure complements the formal computational complexities derived in detail in Sections~\ref{sec.2.1}--\ref{sec.2.2}, and  comprises also the pre-computation costs, which, however, do not need to be paid once again when predicting at multiple locations. For instance, in our implementation of the VB strategy in   \url{https://github.com/danieledurante/PredProbitGP}, the overall runtime in seconds for predicting at $100$ locations almost coincides with the one reported in Table \ref{tab:T1}  for a single prediction.

As illustrated in the tutorial implementation of all the methods analyzed --- which is available at \url{https://github.com/danieledurante/PredProbitGP/blob/main/Tutorial.md} --- Monte Carlo inference under STAN \citep{hoff_2014} relies on the highly-optimized state-of-the-art \texttt{R} package \texttt{rstan} applied to model \eqref{eq1} for obtaining posterior samples from $\bof(\bX)$, which are then used to compute the predictive probabilities at the test locations via ordinary kriging. Such evaluations rely on $10{,}000$ MCMC samples after a burnin of $10{,}000$, setting the true $\balpha=(\sqrt{30},\sqrt{30})$. In evaluating the performance of minimax tilting (TN) \citep{botev_2017}, we  compute the numerator and the denominator in \eqref{eq4} separately via the \texttt{R} package  \texttt{TruncatedNormal}, using the default settings. Equation \eqref{eq4} is also evaluated under the TLR method presented in Section \ref{sec.2.1} and summarized in Algorithm \ref{algo:1}, which can be implemented via simple adaptations of the \texttt{R} package  \texttt{tlrmvnmvt} \citep{cao2020}. In implementing this routine, we set the block size to $n^{1/2}$, the truncation level to $10^{-4}$ and $R=20{,}000$. To evaluate the predictive probabilities under TN and TLR, we avoid setting $\balpha$ at the true values $(\sqrt{30},\sqrt{30})$, but instead estimate these two GP hyperparameters via the grid search  discussed in Section~\ref{sec.2.1}, that evaluates the marginal likelihood in \eqref{eq3} on a $10 \times 10$ grid in $[\sqrt{15}; \sqrt{45}] \times [\sqrt{15}; \sqrt{45}] \in \mathbb{R}^2$ leveraging  the \texttt{R} packages \texttt{TruncatedNormal} and  \texttt{tlrmvnmvt}, for TN and TLR, respectively. Results are comparable, although TLR requires substantially lower runtimes. The estimate of $\balpha$ provided by  \texttt{tlrmvnmvt} is also used in the implementation of the VB presented in Section \ref{sec.2.2} and summarized in Algorithm \ref{algo:2}. Also in this case we consider $R=20{,}000$ Monte Carlo samples to evaluate \eqref{eq10} via \eqref{eq13}. Such values are generated from the optimal univariate truncated normal approximating densities produced by the CAVI in Algorithm \ref{algo:2}, which can be implemented via minor adaptations of the code in the GitHub repository \texttt{Probit-PFMVB} \citep{fasano2019asymptotically}. 

As clarified in Table \ref{tab:T1}, the methods proposed in Sections~\ref{sec.2.1} and \ref{sec.2.2} notably reduce the runtimes relative to state-of-the-art competitors, thus making prediction under probit GP computationally feasible in those high-dimensional settings that often arise in various applications. According to the MSEs reported in Table \ref{tab:T1}, such a notable reduction in runtimes under TLR and VB  is crucially obtained at almost no costs in terms of  accuracy in the estimation of the predictive probabilities, when compared to relevant competitors relying on MCMC samples  from the exact posterior (STAN) or on accurate evaluation of multivariate Gaussian cumulative distribution functions (TN). The runtimes of  TLR and VB  are also coherent with the associated $\mathcal{O}(n^{5/3}+Rn^{3/2})$ and  $\mathcal{O}(n^{3}+Rn)$ computational costs discussed in Sections \ref{sec.2.1}--\ref{sec.2.2}, which make VB more competitive in small-to-moderate dimensions, and TLR more suitable in much higher dimensions due to the reduction of the cubic pre-computation cost. All computations were run on a 3.4 GHz Intel Core i5 CPU workstation, without multithreading.

\section{Saudi Arabia Windspeed Application}
\label{sec.4}

\begin{figure*}[t]
	\centering
		\includegraphics[scale=.59]{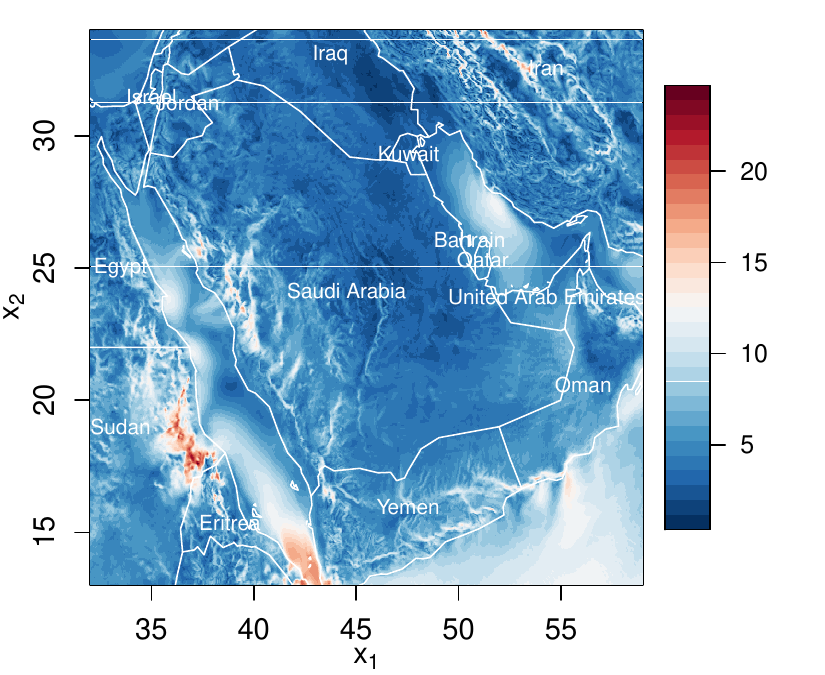}
		\includegraphics[scale=.59]{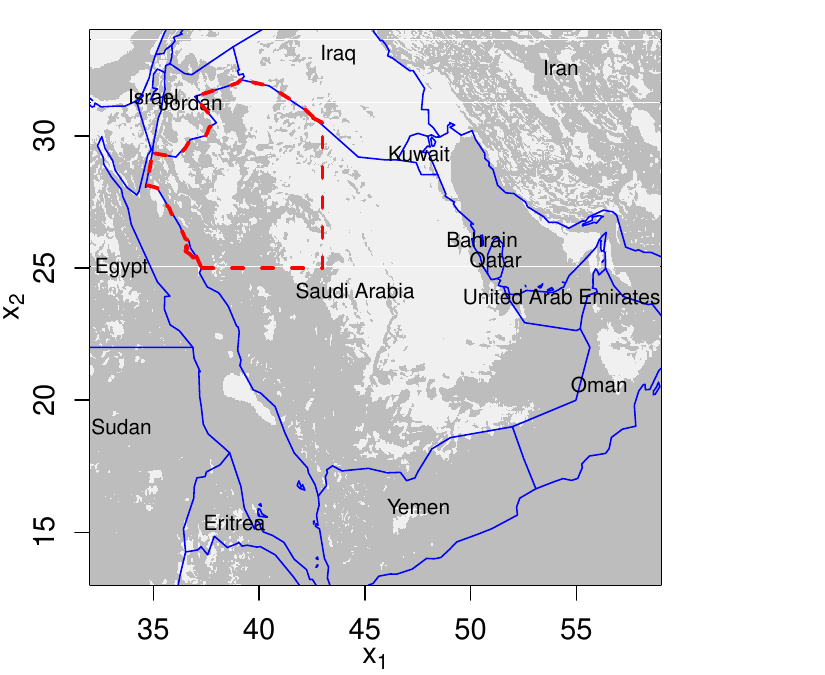}	
		\vspace{-5pt}	
	\caption{{\footnotesize{Heatmaps representing the windspeed at $140$ meters high (left) and a binary version $y$ of this measure defining whether the local windspeed is sufficiently high for energy production (dark gray: YES; light gray: NO) based on the 4m/s threshold (right) on Jan 21st, 2014. The dashed area denotes the spatial region that is used for modeling and prediction.}}}
	\label{fig:F2}
\end{figure*}

We conclude by applying the methods developed in Sections \ref{sec.2.1} and \ref{sec.2.2} to a real-world environmental application aimed at modeling whether the local windspeed exceeds a pre-specified working threshold for energy production in a given region of interest in Saudi Arabia. 
Wind turbines for generating electricity typically have two windspeed thresholds, of which the lower controls when the blades of the turbine start to be in motion and the higher indicates if the turbine should be switched off to avoid strong-wind damage. Here, the binary response $y_i \in \{0;1\}$ measures whether the windspeed at the $i$-th location exceeds the lower threshold, thus allowing production of wind power, which is referred to as the working threshold of wind turbines. 
This important application is motivated by the growing domestic energy consumption in Saudi Arabia and by the attempt to reduce the reliance on fossil fuels, thereby leading to an increasing interest on renewable energy sources, including wind \citep[][]{shaahid2014,chen2018,tagle2019,giani2020}. The effective exploitation of such resources and the careful management of the energy stations require careful modeling and prediction at a fine spatial resolution of whether the local windspeed exceeds or not a given threshold for energy production. As discussed in the following, this fine grid of observations commonly produces a sample size around tens of thousands units. This makes state-of-the-art algorithms for probit GP computationally unfeasible, thus motivating  the use of our scalable solutions in Sections \ref{sec.2.1}--\ref{sec.2.2}. 

\begin{figure*}[t]
	\centering
		\includegraphics[scale=.58]{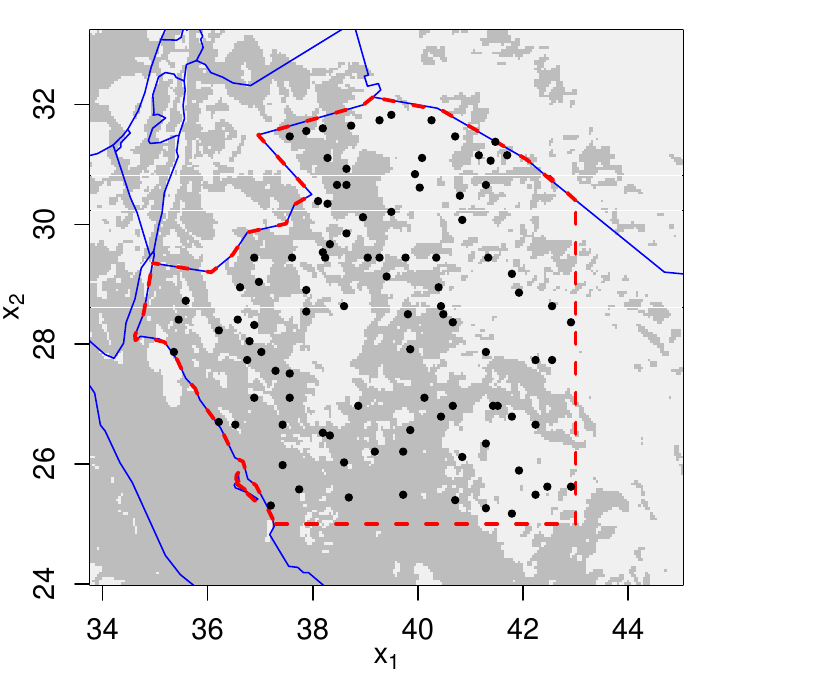}
		\includegraphics[scale=.58]{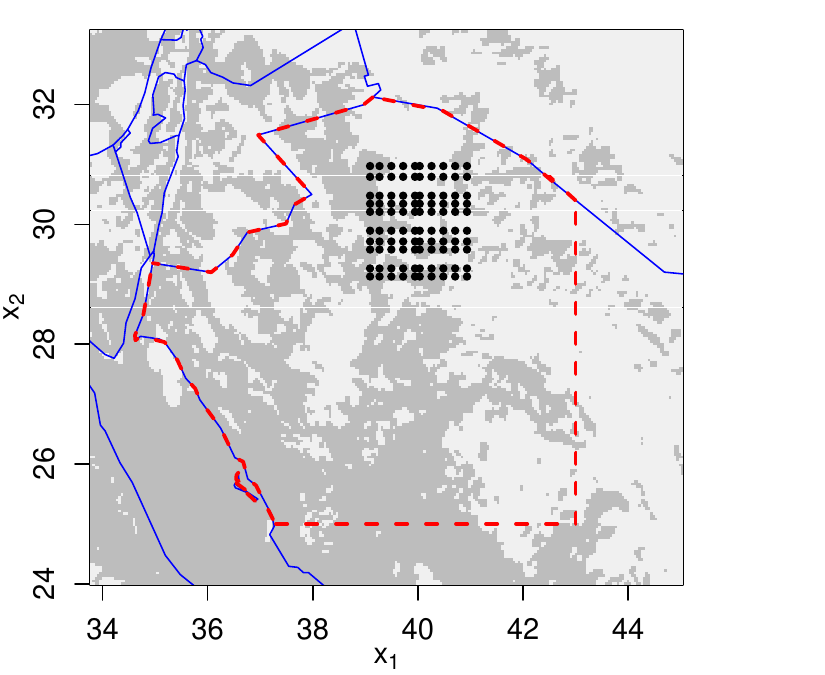}	
				\vspace{-2pt}		
	\caption{{\footnotesize{For the the spatial region used in modeling and prediction, heatmaps defining whether the local windspeed is sufficiently high for energy production (dark gray: YES; light gray: NO) based on the 4m/s threshold on Jan 21st, 2014. Black circles denote the 100 test locations distributed randomly (left) and on a grid (right), used for prediction.}}}
	\label{fig:F3}
\end{figure*}

The windspeed dataset considered in this article is produced by the Weather Research and Forecasting (WRF) model \citep{yip2018statistical}, which constructs the weather system through partial differential equations on the mesoscale and demands strong computation capacity to serve meteorological applications \citep{skamarock2008}. The time resolution of our data is daily and we use windspeed over the region of north-west Saudi Arabia on January 21st, 2014 for modeling and out-of-sample prediction. Such a region covers the wind farm at Dumat Al Jandal, which is the first wind farm in Saudi Arabia and currently under construction, as well as the future smart city of NEOM, a strategic component of the Saudi 2030 Vision, where wind power is expected to be a key energy source. Moreover, the windspeed on January 21st, 2014 has high variability across this region, which makes the out-of-sample prediction task much more challenging. As shown in Figures \ref{fig:F2} and \ref{fig:F3} the region under analysis is obtained by intersecting the Saudi Arabia territorial map with the rectangle ranging from \textsc{e}$34^\circ 30^\prime$ to \textsc{e}$43^\circ$ and from \textsc{n}$25^\circ$ to \textsc{n}$32^\circ$. Within this region we consider a fine grid of $n = 9{,}036$ equally-spaced locations $\bx_i=(x_{i1},x_{i2})^{\intercal}=(\texttt{long}_i,\texttt{lat}_i)^{\intercal}$ at which we monitor whether the windspeed is either above $(y_i=1)$ or below $(y_i=0)$ the working threshold of wind turbines for each $i=1, \ldots, 9{,}036$. Following \citet{chen2018}, such a threshold is set at 4 m/s, leading to a balanced dataset with $51\%$ `1' responses, and $49\%$ observed `0's. Similar to Section \ref{sec.3}, we monitor predictive performance at $100$ out-of-sample locations displayed in Figure \ref{fig:F3}, which are distributed randomly, and on a grid centered at the Dumat Al Jandal wind farm.

Motivated by the results in the simulation study in Section~\ref{sec.3}, we consider a probit GP with zero mean  and squared exponential covariance kernel 
\begin{equation*}
K(\bx,\bx')=\exp\{-[\alpha^2_1(x_1{-}x'_1)^2 +\alpha^2_2(x_2{-}x'_2)^2]\},
\end{equation*}
where $\balpha=(\alpha_1,\alpha_2)$ is estimated via a grid maximization of the marginal likelihood in \eqref{eq3} evaluated via the  \texttt{tlrmvnmvt} package on a $20 \times 20$ grid of values in $[1;\sqrt{30}] \times [1;\sqrt{30}]$.  The estimated  $\balpha$ is  $(3.59, 4.77)$, which interestingly implies a similarly rapid decay in correlation across the two spatial directions. This result is consistent with the abrupt changes of the binary responses.  Recalling the results in Table \ref{tab:T1}, calculation of the predictive probabilities is only performed under the methods presented in Sections \ref{sec.2.1} (TLR) and \ref{sec.2.2} (VB) since STAN and TN would be computationally impractical in such a high-dimensional setting with $n = 9{,}036$. Although this issue could be circumvented via subsampling, such a procedure is suboptimal since it reduces the sample size $n$ and, as  a consequence, it yields less accurate estimates of the predictive probabilities with higher MSE; see also Table \ref{tab:T1}. In implementing both methods, we set $\balpha=(3.59, 4.77)$ and consider the same settings as in the simulation study in Section \ref{sec.3}, thus obtaining runtimes that are comparable to those discussed in Section \ref{sec.3} for the scenario with $n=10{,}000$. Out-of-sample predictive performance measured via the area under the ROC curve (AUC) is similarly accurate for both methods. In particular, the AUCs for the random and grid test scenarios are above $0.9$ under both TLR and VB. This confirms the accuracy gains that can be obtained by the development of increasingly scalable strategies which can be effectively applied to larger samples sizes.

\section{Discussion}
\label{sec.5}
This article provides novel expressions for the predictive probabilities under probit models with GP priors, relying either on multivariate Gaussian cumulative distribution functions or on functionals of multivariate truncated normals, and proposes scalable computational strategies to evaluate such quantities in common high-dimensional settings, thus covering an important gap in the literature. As highlighted in the simulations studies in Section \ref{sec.3}, such computational gains are notable and do not sacrifice accuracy. This allows effective exploitation of the full information in the observed data to improve predictive accuracy, even in computationally challenging applications, such as the  windspeed study  in Section \ref{sec.4}, where the high sample size affects the practical feasibility of available state-of-the-art solutions.

The above results open up several avenues for future research. A relevant direction is to address the possible computational bottlenecks of the proposed grid search for hyperparameter tuning in settings when the dimension of $\balpha$ is large. This issue arises in high-dimensional predictor domains when considering, for example,  automatic relevance determination (ARD) kernels that assign a different scaling hyperparameter for each predictor  \citep[e.g.,][Ch.\ 4.2 and 5.1]{ras_2006}. Direct application of the proposed grid search would be computationally challenging in this high-dimensional hyperparameter space as it would require an excessive number of evaluations of the marginal likelihood, unless some assumptions are made on the kernel function to reduce the number of hyperparameters. Although these simplifications are sometimes made in practice  \citep[e.g.,][]{kuss2005,nic2008}, it would be still desirable to develop scalable tuning strategies in high-dimensional hyperparameter spaces. A promising direction to address this goal is to combine our improved strategy for the evaluation of the marginal likelihood in Section~\ref{sec.2.1} with state-of-the-art machine learning algorithms for high-dimensional hyperparameter tuning that require a low number of evaluations of the objective function \citep[e.g.,][]{Bergstra2011,Snoek2012,Klein2017}.

Another area of interest is direct estimation and uncertainty quantification on linear relationships among the response and predictors, when included within the GP  mean function  via $\bx^{\intercal} \bbeta$. Although such a goal departs from the predictive focus of this article, it shall be noticed that the posterior distribution $p(\bbeta {\mid} \by)$ of the regression coefficients can be derived in closed form when considering Gaussian priors for $\bbeta$. In particular, note that when $f(\bx)$ is a GP with mean function $m(\bx)=\bx^{\intercal}\bbeta$ and covariance kernel $K(\bx,\bx')$, then, leveraging standard GP properties, it holds that 
\begin{equation*}
\bof(\bX)=\left(f(\bx_1), \ldots,f(\bx_n)\right)^{\intercal}=\bX\bbeta+\bar{\bof}(\bX)=\bar{\bX}\boeta,
\end{equation*}
where $\bar{\bX}=(\bX, \bI_n)$, $\boeta=(\bbeta^{\intercal},\bar{\bof}(\bX)^{\intercal})^{\intercal}$, and $\bar{\bof}(\bX) \sim \mbox{N}_n({\bf 0}, \bOmega)$, with  $\bOmega$ defined as in~\eqref{eq1}. Hence, letting $\bbeta \sim \mbox{N}_{q}({\bf 0}, \delta^2 \bI_q)$, as in Section~\ref{sec.1},  it follows that $\boeta \sim \mbox{N}_{q+n}({\bf 0}, \bOmega_{\boeta})$, where $\bOmega_{\boeta}$ is a $(q+n) \times (q+n)$ block-diagonal covariance matrix with blocks $\bOmega_{\boeta[1,1]}= \delta^2 \bI_q$ and $\bOmega_{{\boeta}[2,2]}= \bOmega$. Recalling  Sections~\ref{sec.1} and \ref{sec.2}, this multivariate Gaussian prior, when combined with the probit likelihood via the Bayes rule, yields the posterior distribution 
\begin{equation*}
p(\boeta \mid \by) \propto p(\boeta)p(\by \mid \boeta)= \phi_{q+n}(\boeta; \bOmega_{\boeta})\Phi_n(\bD_{\boeta}\boeta; \bI_n), 
\end{equation*}
with $\bD_{\boeta}=\mbox{diag}(2y_1-1, \ldots, 2y_n-1) \bar{\bX}$, whose kernel can be shown to coincide with that of the unified skew-normal variable 
\begin{equation*}
\mbox{SUN}_{{q+n},n}({\bf 0},\bOmega_{\boeta},\bar{\bOmega}_{\boeta}\bomega_{\boeta}\bD_{\boeta}^{\intercal}\bs_{\boeta}^{-1},{\bf 0}, \bs_{\boeta}^{-1}(\bD_{\boeta}\bOmega_{\boeta}\bD_{\boeta}^{\intercal}+\bI_n)\bs_{\boeta}^{-1}),
\end{equation*}
 with $\bs_{\boeta}=[(\bD_{\boeta}\bOmega_{\boeta}\bD_{\boeta}^{\intercal}{+}\bI_n)\odot \bI_n]^{1/2}$, $\bar{\bOmega}_{\boeta}=\bomega_{\boeta}^{-1}\bOmega_{\boeta}\bomega_{\boeta}^{-1}$ and $\bomega_{\boeta}=(\bOmega_{\boeta} \odot \bI_{q+n})^{1/2}$,  leveraging the recent conjugacy results in Theorem 1 of \citet{Durante2018}. Notably, such a class of distributions is closed under marginalization \citep{arellano_2006,azzalini_2013}, meaning that also the posterior distribution $p(\bbeta \mid \by)$  for $\bbeta$ --- which corresponds to the first $q$ entries in $\boeta$ --- is  unified skew-normal with parameters that can be directly obtained from those of the joint SUN posterior for  $\boeta$ via simple linear algebra operations; see \citet[][Ch. 7.1.2]{azzalini_2013} for details. This result facilitates estimation and uncertainty quantification for $\bbeta$, when this is of interest, leveraging the functionals of the associated closed-form SUN posterior \citep[][]{Durante2018}. 

Finally, it is worth emphasizing that the methods developed in Section \ref{sec.2} can be naturally adapted to any probit model with a multivariate Gaussian prior for the linear predictor. Relevant examples include classical Bayesian probit regression, multivariate probit models \citep[e.g.,][]{chib_1998,fasano2019closed} and general additive representations relying on basis expansions. Extensions to categorical response data under a multinomial probit GP model or to more general SUN priors can also be explored by leveraging results in \citet{Durante2018}, \citet{fasano2020} and \citet{benavoli2020}.

{\footnotesize{\section*{\small Acknowledgments}
This publication is based upon work supported by the King Abdullah University of Science and Technology (KAUST) Office of Sponsored Research (OSR) under Award No: OSR-2018-CRG7-3742. }}

\appendix

\section{Appendix: Proof of Theoretical Results}
To prove Propositions \ref{prop1}--\ref{prop2} let us first state the following Lemma.
\begin{Lemma}[Lemma 7.1 in \citet{azzalini_2013}]
If $\bU \sim \mbox{\normalfont N}_p(\bzero, \bSigma)$ then $\mathbb{E}[\Phi_q(\bH^{\intercal}\bU +\bk; \bPsi)]=\Phi_q(\bk; \bPsi+\bH^{\intercal}\bSigma\bH)$, for any choice of the vector $\bk \in \Bbb{R}^q$, the $p \times q$ matrix $\bH$ and the $q \times q$ symmetric positive--definite matrix $\bPsi$.
\label{lemma1}
\end{Lemma}

Combining the closure under conditioning property of multivariate Gaussians with the above result --- whose proof can be found in \citet{azzalini_2013} --- the proof of  Propositions \ref{prop1}--\ref{prop2} can be obtained via simple derivations described below.

\begin{proof}[Proof of Proposition \ref{prop1}] To prove Proposition \ref{prop1}, first to notice that by application of the Bayes rule 
\begin{equation*}
\pr(y_{n + 1}=1 \mid \by) = p(y_{n + 1} = 1, \by) / p(\by).
\end{equation*}
 Hence, it suffices to show that 
 \begin{equation*}
 \begin{split}
&p(y_{n + 1} = 1, \by)=\Phi_{n+1}(\bD^*\bxi^*; \bI_{n+1}+\bD^*\bOmega^*\bD^{*\intercal}),\\
&p(\by)= \Phi_n(\bD\bxi; \bI_n+\bD\bOmega\bD^{\intercal}).
\end{split}
\end{equation*}
Recalling our discussion in Section \ref{sec.2.1}, $p(\by)$ is the marginal likelihood for the observed data and can be expressed as 
 \begin{equation*}
 \begin{split}
p(\by)&={\int} \Phi_n(\bD\bof(\bX); \bI_n) \phi_n(\bof(\bX)-\bxi; \bOmega)\mbox{d} \bof(\bX)\\
&=\mathbb{E}[\Phi_n(\bD(\bof(\bX)-\bxi) +\bD\bxi; \bI_n)],
\end{split}
\end{equation*}
 where $(\bof(\bX)-\bxi) \sim \mbox{N}_n(\bzero, \bOmega)$. Hence, by applying Lemma \ref{lemma1} to this expectation, we obtain 
 \begin{equation*}
 \mathbb{E}[\Phi_n(\bD(\bof(\bX)-\bxi) +\bD\bxi; \bI_n)]= \Phi_n(\bD\bxi; \bI_n+\bD\bOmega\bD^{\intercal}),
\end{equation*}
Such a result also clarifies equation \eqref{eq3}. The proof of equation $p(y_{n + 1} = 1, \by)=\Phi_{n+1}(\bD^*\bxi^*; \bI_{n+1}+\bD^*\bOmega^*\bD^{*\intercal})$ proceeds in a similar manner, after noticing that 
\begin{equation*}
 \begin{split}
&p(y_{n + 1} = 1, \by)\\
&{=}{\int} \Phi(f(\bx_{n+1})) \Phi_n(\bD\bof(\bX); \bI_n) \phi_{n+1}(\bof^*(\bX)-\bxi^*; \bOmega^*)\mbox{d} \bof^*(\bX)\\
&{=}{\int} \Phi_{n+1}(\bD^*\bof^*(\bX); \bI_{n+1}) \phi_{n+1}(\bof^*(\bX)-\bxi^*; \bOmega^*)\mbox{d} \bof^*(\bX)\\
&{=}\ \mathbb{E}[\Phi_{n+1}(\bD^*(\bof^*(\bX)-\bxi^*) +\bD^*\bxi^*; \bI_{n+1})],
\end{split}
\end{equation*}
where $\bof^*(\bX)-\bxi^*=[(\bof(\bX)^{\intercal},f(\bx_{n+1}))^{\intercal}-\bxi^*] \sim \mbox{N}_{n+1}(\bzero, \bOmega^*)$, with $\bxi^*, \bOmega^*$ and $\bD^*$ defined as in Proposition \ref{prop1}.
\end{proof}

\begin{proof}[Proof of Proposition \ref{prop2}] Recalling the results discussed in Section \ref{sec.2.2}, the predictive probability $\pr(y_{n+1}=1 \mid \by)$ can be defined as $\mathbb{E}_{f(\bx_{n+1}) \mid \by}[ \Phi(f(\bx_{n+1}))]$, with $p(f(\bx_{n+1}) \mid \by)$ being the marginal in the joint conditional density $p(f(\bx_{n+1}),\bof(\bX), \bz \mid \by)$ which factorizes as $p(f(\bx_{n+1}) \mid \bof(\bX)) p(\bof(\bX) \mid \bz) p(\bz \mid \by)$. Hence, by the law of the total expectation, we have that 
\begin{equation*}
 \begin{split}
&\pr(y_{n{+}1}= 1 \mid \by)\\
&\qquad = \mathbb{E}_{\bz \mid \by}[\mathbb{E}_{\bof(\bX) \mid \bz}(\mathbb{E}_{f(\bx_{n+1}) \mid \bof(\bX)}[\Phi(f(\bx_{n+1}))])].
\end{split}
\end{equation*}
Since $(f(\bx_{n+1}) \mid \bof(\bX)) \sim \mbox{N}(\mu_{x_{n+1}}+\bH_{x_{n+1}}\bof(\bX), \sigma^2_{x_{n+1}})$ by \eqref{eq8}, we can leverage Lemma \ref{lemma1} above to obtain 
\begin{equation*}
 \begin{split}
&\mathbb{E}_{f(\bx_{n+1}) \mid \bof(\bX)}[\Phi(f(\bx_{n+1}))]\\
& \qquad =\Phi(\mu_{x_{n+1}}+\bH_{x_{n+1}}\bof(\bX);1+\sigma_{x_{n+1}}^2 ).
\end{split}
\end{equation*}
To conclude the proof note that, by \eqref{eq9},  we have $(\bof(\bX) \mid \bz) \sim \mbox{N}_n(\bmu_{\bX}+\bSigma_{\bX}\bz, \bSigma_{\bX})$. Therefore, further application of Lemma \ref{lemma1} yields 
\begin{equation*}
 \begin{split}
&\mathbb{E}_{\bof(\bX) \mid \bz}[\Phi(\mu_{x_{n+1}}+\bH_{x_{n+1}}\bof(\bX);1+\sigma_{x_{n+1}}^2 )]\\
& \qquad  =\Phi(\mu_{x_{n+1}}+ \bH_{x_{n+1}} (\bmu_{\bX}+ \bSigma_{\bX}\bz);\\
& \qquad  \qquad   \qquad  \qquad 1+\sigma_{x_{n+1}}^2+ \bH_{x_{n+1}} \bSigma_{\bX}\bH^{\intercal}_{x_{n+1}})
\end{split}
\end{equation*}
as in Proposition \ref{prop2}.
\end{proof}

\begingroup
\fontsize{11pt}{12pt}\selectfont

\endgroup

\end{multicols}

\end{document}